%% file: fetl.tex
\documentclass[sigconf, nonacm]{acmart}
\usepackage{amsmath,amsfonts}
\usepackage{algorithmic}
\usepackage{graphicx}
\usepackage{textcomp}
\usepackage{pgfplots}
\usepackage{xcolor}
\usepackage{subfigure}
\usepackage{enumitem}

\newtheorem{example}{Example}[section]

\newcommand{\Paragraph}[1]{~\vspace*{-0.8\baselineskip}\\{\textbf{ #1.}}}

\newenvironment{pkl}{%
	\begin{itemize}[leftmargin=*, topsep=3pt]%
		\setlength\itemsep{-0.5\parskip}%
		\setlength\parsep{0in}%
	}{%
\end{itemize}}

\newenvironment{enu}{%
	\begin{enumerate}[leftmargin=*]%
		\setlength\itemsep{-\parskip}%
		\setlength\parsep{0in}%
	}{%
		\vspace{-\topsep}%
\end{enumerate}}

\newcommand\vldbdoi{XX.XX/XXX.XX}

\newcommand\vldbavailabilityurl{URL_TO_YOUR_ARTIFACTS}
\begin{document}
\title{Accelerating Fresh Data Exploration with Fluid ETL Pipelines}

%%
%% The "author" command and its associated commands are used to define the authors and their affiliations.
\author{Maxwell Norfolk}
\affiliation{%
    \institution{The Pennsylvania State University}
}
\email{mnorfolk@psu.edu}

\author{Dong Xie}
\affiliation{%
    \institution{The Pennsylvania State University}
}
\email{dongx@psu.edu}

%%
%% The abstract is a short summary of the work to be presented in the
%% article.
\input{abstract}

\maketitle
%
%%%% do not modify the following VLDB block %%
%%%% VLDB block start %%%
%\pagestyle{\vldbpagestyle}
%\begingroup\small\noindent\raggedright\textbf{PVLDB Reference Format:}\\
%\vldbauthors. \vldbtitle. PVLDB, \vldbvolume(\vldbissue): \vldbpages, \vldbyear.\\
%\href{https://doi.org/\vldbdoi}{doi:\vldbdoi}
%\endgroup
%\begingroup
%\renewcommand\thefootnote{}\footnote{\noindent
%This work is licensed under the Creative Commons BY-NC-ND 4.0 International License. Visit \url{https://creativecommons.org/licenses/by-nc-nd/4.0/} to view a copy of this license. For any use beyond those covered by this license, obtain permission by emailing \href{mailto:info@vldb.org}{info@vldb.org}. Copyright is held by the owner/author(s). Publication rights licensed to the VLDB Endowment. \\
%\raggedright Proceedings of the VLDB Endowment, Vol. \vldbvolume, No. \vldbissue\ %
%ISSN 2150-8097. \\
%\href{https://doi.org/\vldbdoi}{doi:\vldbdoi} \\
%}\addtocounter{footnote}{-1}\endgroup
%%%% VLDB block end %%%
%
%%%% do not modify the following VLDB block %%
%%%% VLDB block start %%%
%\ifdefempty{\vldbavailabilityurl}{}{
%\vspace{.3cm}
%\begingroup\small\noindent\raggedright\textbf{PVLDB Artifact Availability:}\\
%The source code, data, and/or other artifacts have been made available at \url{\vldbavailabilityurl}.
%\endgroup
%}

    %\begin{IEEEkeywords}
    %    component, formatting, style, styling, insert.
    %\end{IEEEkeywords}

    \input{intro}
    \input{tradeoff}
    \input{ingestion}
    \input{query}
    \input{optimize}
    \input{conc}
%%
%% The next two lines define the bibliography style to be used, and
%% the bibliography file.
\bibliographystyle{ACM-Reference-Format}
    \bibliography{fetl}

\end{document}

%% file: abstract.tex
\begin{abstract}
Recently, we have seen an increasing need for fresh data exploration, where data analysts seek to explore the main characteristics or detect anomalies of data being actively collected. In addition to the common challenges in classic data exploration, such as a lack of prior knowledge about the data or the analysis goal, fresh data exploration also demands an ingestion system with sufficient throughput to keep up with rapid data accumulation. However, leveraging traditional Extract-Transform-Load (ETL) pipelines to achieve low query latency can still be extremely resource-intensive as they must conduct an excessive amount of data preprocessing routines (DPRs) (e.g., parsing and indexing) to cover unpredictable data characteristics and analysis goals. To overcome this challenge, we seek to approach it from a different angle: leveraging occasional idle system capacity or cheap preemptive resources (e.g., Amazon Spot Instance) during ingestion. In particular, we introduce a new type of data ingestion system called fluid ETL pipelines, which allow users to start/stop arbitrary DPRs on demand without blocking data ingestion. With fluid ETL pipelines, users can start potentially useful DPRs to accelerate future exploration queries whenever idle/cheap resources are available. Moreover, users can dynamically change which DPRs to run with limited resources to adapt to users' evolving interests. We conducted experiments on a real-world dataset and verified that our vision is viable. The introduction of fluid ETL pipelines also raises new challenges in handling essential tasks, such as ad-hoc query processing, DPR generation, and DPR management. In this paper, we discuss open research challenges in detail and outline potential directions for addressing them.
\end{abstract}

%% file: intro.tex
\section{Introduction}
\label{sec:intro}
Recently, there has been increasing interest in fresh data exploration~\cite{tsm-bench,mach-demo}, where data analysts seek to extract timely insights from \emph{actively collected data}. In such applications, users usually have \emph{little to no prior knowledge} of incoming data or future analysis goals. Thus, they have to devise ad-hoc queries on the spot and refine them in subsequent iterations.
For instance, during a suspected cyberattack, cloud administrators attempt to diagnose the cause and influence
 of an application's spiking response time by inspecting real-time data center logs.
They may start by examining the latency distribution of related services and the average CPU usage in the past hour. Based on initial results, they may explore other metrics (e.g., service category, memory utilization) further in the past, or
delve deeper into detailed statistics at finer granularity.
 In addition, unlike the classic data exploration that concerns static or infrequently updated datasets, fresh data exploration operates on active data streams to discover patterns and understand events as they \emph{manifest in real-time}. Thus, to ensure an interactive experience that delivers timely insights about the latest data, fresh data exploration systems face unique challenges in providing high enough ingestion throughput to keep up with rapid data collection while simultaneously maintain low query latency.

Unfortunately, it is still \emph{prohibitively expensive} to deliver an ideal fresh data exploration experience. On the one hand, we can dump incoming data of various formats (e.g., JSON, CSV, raw text) without any preprocessing and process ad-hoc exploration queries directly on the raw data~\cite{mison,raw,nodb}. However, despite achieving the highest possible ingestion throughput, explorative queries often suffer from high latency as excessive data preparation efforts are postponed until query time.
On the other hand, we can leverage Extract-Transform-Load (ETL) pipelines
which execute a set of predefined \textbf{Data Preprocessing Routines (DPRs)}, such as parsing and building indexes, during ingestion.
%\Max{which during ingestion execute a set of predefined \textbf{Data Preprocessing Routines (DPRs)}}, such as parsing and building indexes
By paying these overheads upfront, future queries can leverage \textbf{DPR-built structures} (e.g., parsed out fields, indexes, materialized views, etc.) to reduce query latency effectively. However, each DPR requires dedicated computing resources (in CPU cycles, memory/disk bandwidth, etc.) to ensure that data ingestion keeps up with data collection. Furthermore, the unpredictable nature of fresh data exploration may cause a mismatch between predefined DPRs and later queries, leaving some exploration queries unoptimized and slow. A common workaround~\cite{docdb} is to exhaustively execute all potentially useful DPRs in ETL pipelines to prepare for all possible queries.
Nevertheless, this approach will result in significant and unnecessary resource overhead.
Lastly, classic data exploration techniques like database cracking~\cite{cracking} have limited power in accelerating fresh data exploration workloads since they focus on reorganizing data queried in the past but do not cover any data ingested in the future.

To overcome this challenge, we seek to approach it from a different angle: leveraging occasional idle system capacity or inexpensive preemptive resources to adaptively conduct DPRs that may accelerate future queries. Specifically, as the data collection rate and characteristics change constantly, data ingestion systems must provision enough computational resources to sustain the peak workload. As a result, during off-peak periods, there could be significant amount of resources left idle in the system. Moreover, modern cloud providers also offer their unused capacity (through services like EC2 Spot Instances~\cite{spotinstance}) with a very deep discount (up to 90\%), allowing opportunistic service deployments at a low cost. Another key insight we leverage to accelerate fresh data exploration is that users usually only care about a \emph{\textbf{small}} but \emph{\textbf{changing}} subset of interesting data at a time. Thus, the minimal DPR effort to achieve good query performance is \textbf{vastly smaller} than the combined effort of all potentially useful ones. In light of this, we see the opportunity to accelerate fresh data exploration through carefully investing such resources adaptively towards meaningful efforts (i.e., DPRs) that will eventually accelerate future queries. Meanwhile, the query engine should also fully leverage any existing efforts (such as DPR-built indexes and intermediate results) to improve query performance.

To realize this vision, we introduce a novel type of ingestion system called \textbf{fluid ETL pipelines}, which can proactively adjust its executing DPRs so it can adapt to the availability of idle/cheap resources and users' changing exploration interests. In contrast to traditional ETL pipelines, DPRs executed in a fluid ETL pipeline are not determined upfront and may change on demand during runtime.
Combining fluid ETL pipelines with adaptive DPR management and careful query planning/optimization, we see a viable path of effectively accelerating fresh data exploration workloads through striking a dynamic balance among \emph{ingestion throughput}, \emph{query latency}, and \emph{resource usage}.

In this paper, we define fluid ETL pipelines by describing their core capabilities and providing
concrete implementation plans to build them. Furthermore, we demonstrate that fluid ETL pipelines can bring significant performance benefits for fresh data exploration with low cost through experiments on real-world datasets. We then outline the key components to develop around fluid ETL pipelines towards building a practical fresh data exploration solution. Lastly, we discuss core technical challenges in developing these components and outline potential directions to address them.
%\Max{Lastly, we discuss the core technical challenges in developing these components, and outline potential directions to address them.}

%% file: tradeoff.tex
\section{Introducing Fluid ETL Pipelines}
\label{sec:fetl}

In this section, we define what a fluid ETL pipeline is and provide a concrete plan to build it. Next, we show how fluid ETL pipelines can help accelerate fresh data exploration workloads with idle system capacity through an experiment conducted on a real-world dataset. We then outline the essential components to be built around a fluid ETL pipeline towards a practical fresh data exploration solution. Lastly, we will describe the connection and differences between existing solutions and ours.

\Paragraph{Defining Fluid ETL Pipelines} Similar to traditional ETL pipelines, the goal of a fluid ETL pipeline is to ingest an incoming data stream, selectively extract essential information from raw data, then transform it into another format, build/maintain auxiliary structures, and/or load them into a database system. Generally, we define the term \textbf{Data Preprocessing Routine (DPR)} as a procedure that takes a set of incoming data as input and produces (or maintains) some structures (e.g., indexes, materialized views, etc.) to accelerate future queries. For example, we can define a DPR that takes a single JSON record, parses out a few fields, evaluates a function with extracted values, and inserts the record into an index.
Another DPR example is taking a set of CSV records, extracting all fields for each record, and reformatting them into Parquet~\cite{parquet}. With this concept, a traditional ETL pipeline can be viewed as an ingestion system which continuously applies \textbf{\emph{a predefined set of DPRs}} to incoming records. In contrast, DPRs executed in a fluid ETL pipeline are \textbf{\emph{not determined upfront}} and may change over time based on resource availability and user demands. Specifically, a fluid ETL pipeline should provide at least the following capabilities:

\begin{figure}[t]
	\centering
	\includegraphics[width=3.4in]{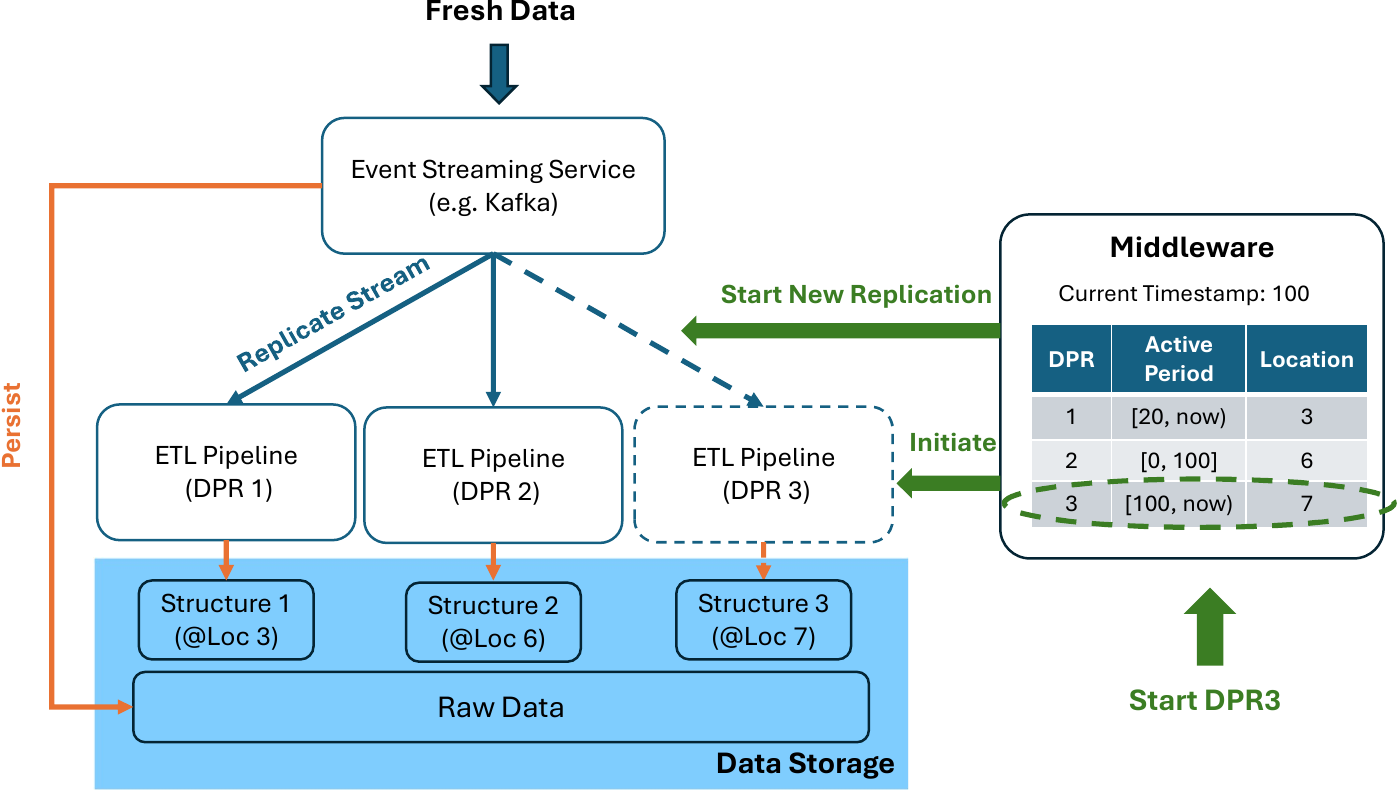}
	\caption{Fluid ETL Pipeline Implementation Plan.}
	%\vspace{-2mm}
	\label{fig:fetl-impl}
	\vspace{-5mm}
\end{figure}

\begin{enu}
	\item Users must be able to start/stop arbitrary DPRs at any time without blocking ingestion. To ensure this, fluid ETL pipelines only apply a DPR to records ingested when it is currently active.
	\item Fluid ETL pipelines must remember active time periods of all previous and currently executing DPRs, along with the locations of corresponding structures they built.
	\item Fluid ETL pipelines must persist a raw copy of ingested data to ensure any valid ad-hoc queries can be answered properly.
\end{enu}

As soon as a piece of data is ingested by the fluid ETL pipeline, its raw form and any structures built for it should become immediately visible to the query engine to ensure result freshness. Due to capability (1) listed above, leveraging DPR-built structures to process ad-hoc queries efficiently becomes a delicate problem. Since a DPR will only be applied to data ingested during its active period, the structures it builds or maintains may also cover only part of, rather than all, ingested data. That means, these structures may be \textbf{\emph{partial}} to an ad-hoc query (i.e., do not cover all related data for answering it), making classic query optimizer unable to leverage them. However, as we will discuss in Section~\ref{sec:query}, these partial structures could still provide significant performance benefits with novel query planning and optimization techniques.

\begin{figure*}[t]
	\subfigure[Ingestion CPU Utilization \& Fluid ETL DPR Setup]{\label{fig:fetl-ingest}\includegraphics[width=.48\textwidth]{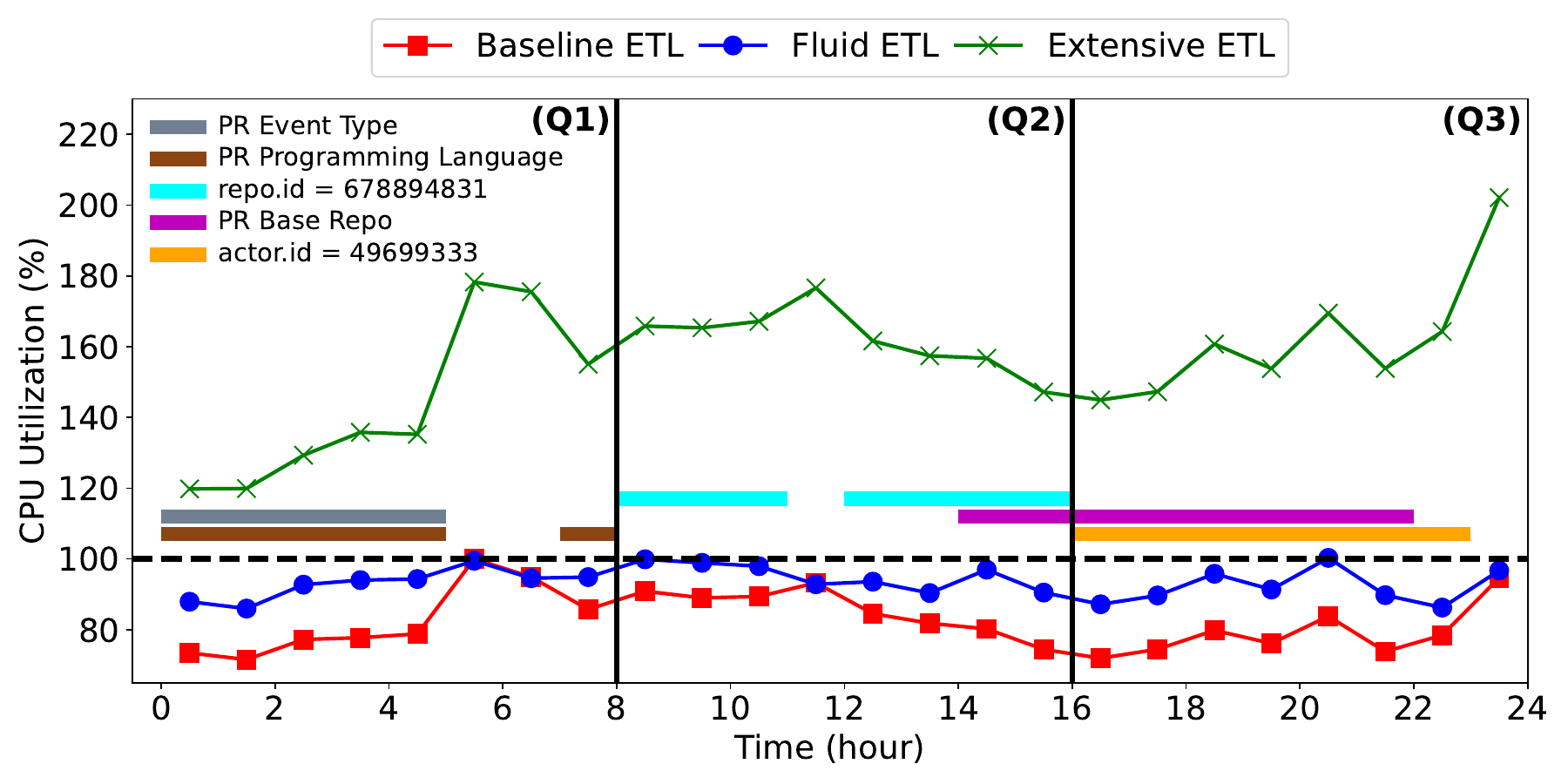}}
	\subfigure[Exploration Query Latency]{\label{fig:fetl-query}\includegraphics[width=.48\textwidth]{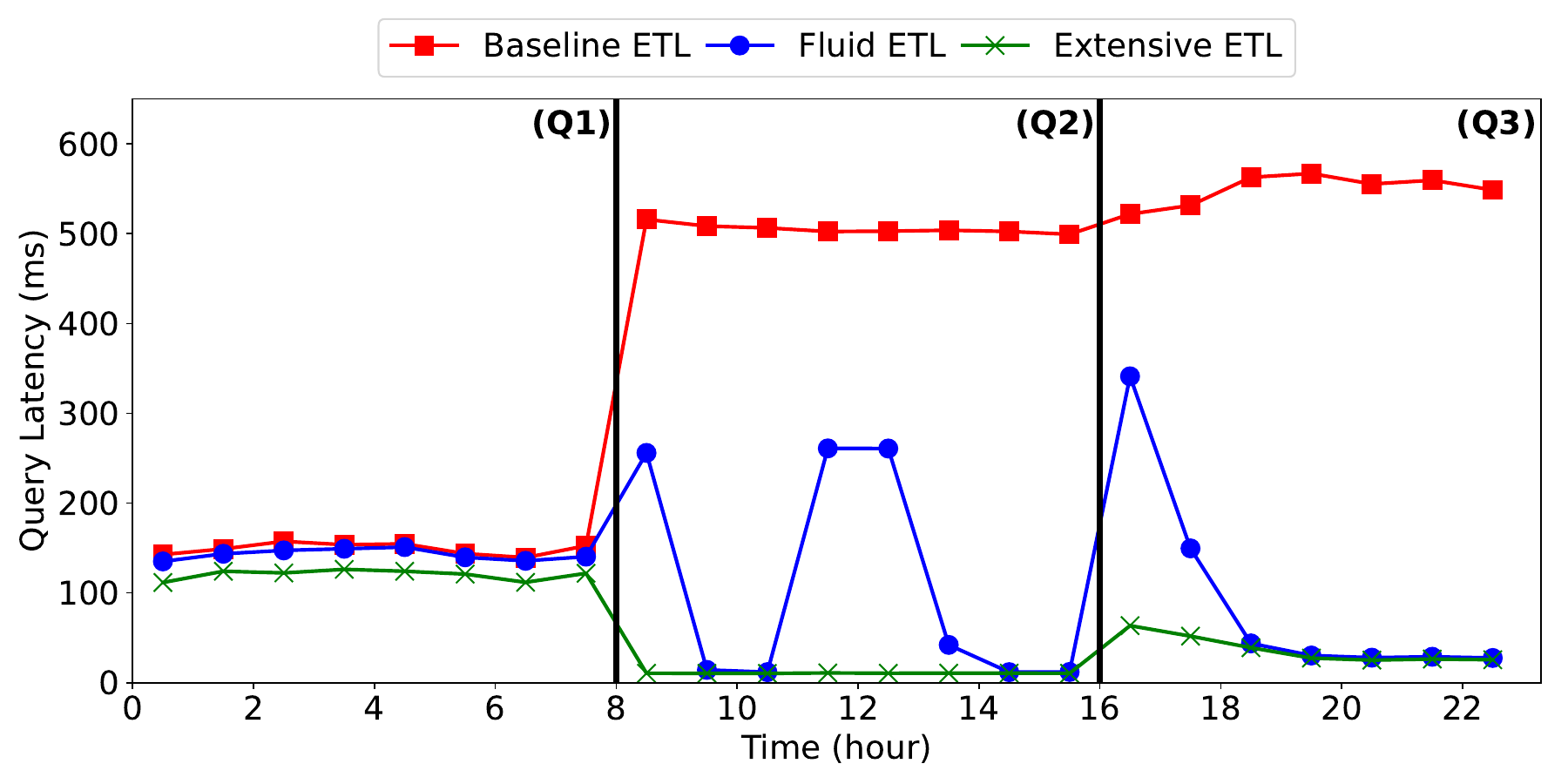}}
	\vspace{-5mm}
	\caption{Ingestion CPU Utilization and Exploration Query Latency for Various ETL Strategies}
	\label{fig:fetl}
	\vspace{-5mm}
\end{figure*}

\Paragraph{Implementing a Fluid ETL Pipeline} To implement a functioning fluid ETL pipeline, we simply have to combine an event streaming system (like Kafka~\cite{kafka}) with multiple traditional ETL pipelines (which could be implemented through Spark~\cite{spark}, Flink~\cite{flink}, etc.). A possible implementation plan is shown in Fig.~\ref{fig:fetl-impl}. Note that a middleware is introduced to modulate the event streaming system and various ETL pipelines. As shown in the figure, when the user wants to introduce a new DPR (DPR 3 in the example) to the system, they just need to implement a traditional ETL pipeline conducting this new DPR and then the middleware will initiate this new ETL pipeline and ask the event streaming system to replicate and route data ingested after that point to the new ETL pipeline. If users turn off a DPR, the middleware would just stop sending later ingested data to the corresponding ETL pipeline and shut it down.
The middleware will also be responsible for tracking the mapping between DPRs to their corresponding structures and the time period they were built.
%which structures were built by which DPR over the data ingested during which period.
% even if we don't keep my sentence, the above sentence has "which" 3 times, so should be changed.
%\Max{which records the DPR was performed on, and the corresponding structures created/maintained during period.}
In principle, this simple design could accommodate arbitrary types and numbers of DPRs. Nevertheless, this is clearly not an efficient design as we have to duplicate the data stream, and multiple ETL pipelines cannot share common efforts. When DPR types are restricted, more efficient designs are possible. For example, our previous work, FishStore~\cite{fishstore}, can serve as a fluid ETL pipeline supporting DPRs that logically organizes records of different properties in a subset hash index.

\vspace{-1mm}
\Paragraph{Performance Benefits from Fluid ETL Pipelines} To demonstrate how much performance benefit fluid ETL pipelines can bring to fresh data exploration, we present a simple application scenario on a real-world stream. In particular, we ingest a day worth of event log from the GitHub Archive dataset~\cite{github} (where events are recorded in JSON documents) and simulate a simple fresh data exploration procedure. The basic experiment setup involves a traditional ETL pipeline that parses out a few common fields (e.g., \texttt{created\_at}, \texttt{type}, etc.) from each event, builds secondary indexes on those fields, and persists a raw copy of each processed record. We provision just enough CPU resources for this ETL pipeline to allow it sustaining real-time data ingestion during the peak hour. Ingested data and constructed indexes resides in memory so we can focus on understanding the behavior of CPU cycle usage. The fresh data exploration scenario we simulate is described as follows:

\begin{example}
	Alice is trying to identify the root cause of an ongoing performance problem for pull request creation at GitHub by examining its active event stream. Initially, she reviews pull requests across all repositories before identifying a particular repository with an unusually high number of recent pull request creations. Next, she reviews all recent events from this repository and observes that a specific account is spamming new events to it. She then examines other repositories where this account is active to determine if it is malicious.
\end{example}

More concretely, we issue the following query sequences against the data stream while it is being ingested: \textbf{(Q1)} find the three repositories with the most pull requests; \textbf{(Q2)} find the top three users of highest activity for a particular repository; \textbf{(Q3)} find the top three repositories where a given user is the most active. All queries are run against the events generated from the last two hours to reflect the online analysis nature. Each query will be executed a few times to track active trends and ensure result stability. %TODO highlight dependencies

We compare the following ETL strategies combined with manually optimized queries on resource utilization during ingestion and exploration query latency.
\begin{pkl}
	\item \textbf{Baseline ETL:} No additional DPRs are introduced on top of the traditional ETL pipeline described in the basic settings above. As a result, only the immediate step \textbf{(Q1)} could benefit from the index built on \texttt{type}. \textbf{(Q2)} and \textbf{(Q3)} cannot leverage any constructed indexes, hence fall back to raw data processing.
	\item \textbf{Fluid ETL:} Set up a fluid ETL pipeline that conducts additional DPRs when there are idle CPU resources and switch what DPRs it runs to adapt to changing queries. More precisely, between hours 1 to 8, the fluid ETL pipeline chooses to run DPRs with idle CPU cycles to build hash indexes on different programming languages and action type for pull request events to prepare for predictions based on \textbf{(Q1)}. As user's query shift to \textbf{(Q2)}, the fluid ETL pipeline adapts its DPR selection to conducted pre-filtering on the queried \texttt{repo.id} while also preparing an index on base repos of pull requests as a backup. Lastly, after \textbf{(Q3)} starts, the fluid ETL pipeline replaced its pre-filtering DPR on \texttt{repo.id} with a filter on the current queried \texttt{actor.id}.
	Note that the fluid ETL pipeline will turn on when there are available CPU resources in the system. Thus, there is no DPR run by the fluid ETL pipeline
	during hours 6, 7 and 12. Additionally, the number of fluid ETL DPRs	varies over time as the amount of idle CPU resources change. The detailed DPR setup for this ETL strategy are shown as color-coded segments in Fig.~\ref{fig:fetl-ingest}.
	\item \textbf{Excessive ETL:} Add six new DPRs to conduct field parsing and index building on various fields (including \texttt{repo.id} and \texttt{actor.id} queried by \textbf{(Q2)} and \textbf{(Q3)}) throughout the ingestion period to the baseline ETL pipeline. Here, we simulate the case where ETL pipelines are configured to do excessive work to cover {\em all foreseeable queries}. All queries can be manually planned to leverage one of the built indexes.
\end{pkl}

As shown in Fig.~\ref{fig:fetl}, we can clearly observe that neither Baseline ETL nor Extensive ETL strategy is desirable in practice. In particular, \textbf{(Q2)} and \textbf{(Q3)} experience significantly longer query latency when only baseline ETL is applied. Extensive ETL requires significantly more CPU resources (up to 2x) than provisioned to deliver great performance on all queries. In contrast, with fluid ETL pipelines adaptively conducting DPRs when there is idle resources in the system, users can enjoy the best of both worlds where the system never demands more CPU cycles than provisioned during ingestion and deliver significantly better query performance ($>$3x in most cases) than Baseline ETL.

\Paragraph{Towards Practical Fresh Data Exploration} Through the experiment above, it is clear that fluid ETL pipelines can bring significant performance benefit for fresh data exploration through leveraging idle/cheap computational resources to run DPRs adaptively. However, several major challenges still remain before we can develop a practical fresh data exploration solution around fluid ETL pipelines. Specifically, we have to handle the following tasks properly:

\begin{enu}
	\item Make predictions on future workloads and generate efficient DPRs to run during ingestion that build structures to accelerate them.
	\item Plan ad-hoc queries carefully to maximize the performance benefits provided by partial structures.
	\item Make dynamic decisions on starting/stopping DPRs in the fluid ETL pipeline to balance ingestion throughput and query latency under resource constraints.
\end{enu}

In the previous experiment, we fulfilled all these tasks through manual implementation and management. However, fulfilling each of these tasks can be very disruptive in the exploration process and requires strong system expertise. Considering most data analysts are not system experts, it is essential to build \textbf{\em automated tools} to help them with these tasks. In the following sections, we will discuss the key technical challenges in developing such tools and outline possible directions to address them.

\begin{figure}
	\centering\includegraphics[width=3.2in]{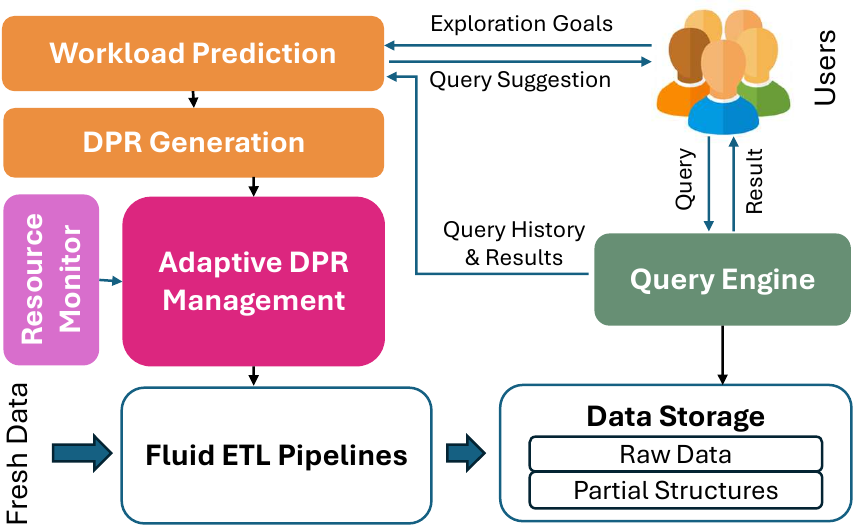}
	\caption{Fresh Data Exploration Solution around Fluid ETL Pipelines}
	\label{fig:solution}
	\vspace{-6mm}
\end{figure}

By integrating fluid ETL pipelines with these automated tools, we envision Fig.~\ref{fig:solution} as a potential system architecture of a practical fresh data exploration solution. In particular, a \emph{resource monitor} will keep track of any idle system resources and/or the amount of available preemptive resource for a given budget. When received an ad-hoc query, the \emph{query engine} will try to fully utilize existing partial structures to maximize their performance benefit. As fresh data exploration proceeds, the system will keep track of all received queries to approximate future workloads. In addition, users can also provide hints on their exploration goals to a \emph{workload prediction} module. Combined with previous queries and their results, this module will predict the most likely next steps in the exploration process to provide recommended queries to users and help improve the quality of future workload approximation. Meanwhile, new DPRs are automatically generated to accelerate predicted workloads. Finally, the \emph{adaptively DPR management module} will choose which DPRs to start/stop in the fluid ETL pipeline based on current availability of idle/cheap resources and predicted workloads.

\Paragraph{Comparison to Existing Solutions} There are apparent differences and connections between existing solutions for fresh data exploration and ours centered around fluid ETL pipelines. Firstly, numerous techniques have been developed to conduct query processing directly on raw data~\cite{mison,sparser,pcsv,raw,nodb}, which are widely used in modern data lakes. These techniques effectively postponed extensive data preparation efforts until query time, resulting in high query latency. With fluid ETL pipelines, although most of the essential data preparation can be predicted and completed upfront, we still need raw data processing techniques to serve when previous DPRs fail to capture the structures required for ad-hoc queries. There are also techniques, such as database cracking~\cite{cracking,scracking} and speculative loading~\cite{specload,nodb}, which gradually extract more information and build structures at query time. Such techniques expect similar query patterns in the future, so structures built by previous queries can accelerate later ones. However, when facing a different query type, these techniques cannot bring any benefits, and must fall back to raw data processing with high query latency. In contrast, our solution will proactively predict future queries and build structures in advance. Nevertheless, we can still leverage these techniques in our solution by treating the by-product of queries as additional partial structures that can be used in future queries. Lastly, classic techniques for accelerating data exploration workloads such as database cracking~\cite{cracking} focus on reorganizing data as they are processed by individual queries. However, it will not apply any transformation on data ingested since the last query, limiting its power to accelerate queries conducted on the freshest data.

As a fluid ETL pipeline effectively contains a changing set of traditional ETL pipelines, all existing techniques to accelerate traditional ETL pipelines~\cite{instant-loading, udp} can also be applied in our solution. There are also systems (e.g., Napa~\cite{napa}) that trade result freshness for robust query performance to maintain reasonable resource usage. However, sacrificing freshness is not always desirable. Even when acceptable, it is hard to determine how much freshness users are willing to sacrifice upfront. Finally, streaming systems~\cite{flink,sparkstreaming} can monitor the results of standing queries over predefined (sliding or tumbling) windows. Nonetheless, in an exploratory setting, users may need to conduct their queries across different time windows or even on historical data, where streaming systems are not helpful.

%% file: ingestion.tex
\section{Generation of Efficient DPRs}
\label{sec:dprgen}

We will first discuss how to predict future workloads and generate efficient DPRs automatically to accelerate them. Similar to classic data exploration applications, users are likely to issue queries with structures and predicates resembling previous ones.
%Hence,
Thus, we can start by predicting future workloads by proportionally mixing query templates seen in the near past.
%In addition, we also acknowledge that new query types could be
Although such approach covers many common cases, new queries types may also be issued naturally, aligning with previous exploration path and a final goal.
With the recent advances in the reasoning capability of Large Language Models (LLMs), we see the opportunity to develop advanced prediction models to forecast the most likely next steps in the exploration procedure. Note that previous query results may play a crucial role here since different past observations may lead to opposite exploration decisions. Thus, in addition to the final goal and query history, the reasoning model must also put query results into context for predicting future queries.

With proper workload prediction, we now get a set of potential queries to accelerate. As we are eventually utilizing structures built/maintained by DPRs (e.g., indexes, materialized views), we should examine the logical plans of potential queries to identify the structures that could be useful to accelerate
them. As multiple queries may benefit from the same structure, we should prioritize structures that help more potential queries. Multiple exploration users may also share some common interests at this moment. To identify these prioritized structures, we could apply classical materialized view and index selection techniques to the predicted workload. For each structure identified, we can leverage existing tools like dbt~\cite{dbt} to describe the transformation processes, generate code to execute them, and utilize a Just-in-Time (JIT) compiler~\cite{llvm-jit} to integrate them into a running fluid ETL pipeline.

\begin{figure}
	\centering
	\includegraphics[width=3in,trim=5mm 6mm 0 0]{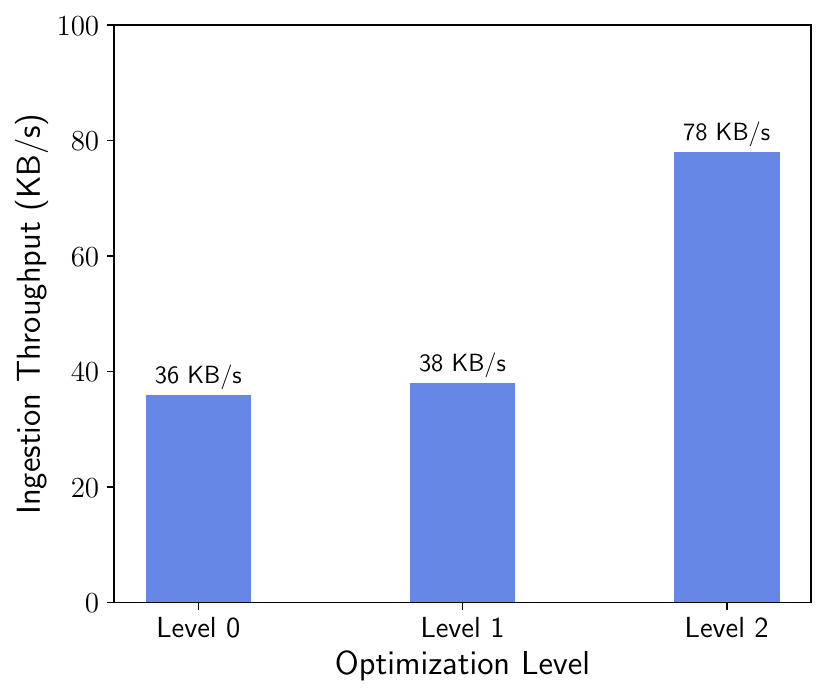}
	\caption{Effect of Sharing Efforts among DPRs}
	\label{fig:fuse}
	\vspace{-5mm}
\end{figure}

Note that the fluid ETL pipeline may run multiple DPRs concurrently to cover a broader exploration frontier. With the simple design we introduced in Section~\ref{sec:fetl}, these DPRs are carried out independently in different ETL pipelines. However, there could be significant logic overlap among these DPRs. For example, we may have two DPRs operating on comments left for issues in the GitHub event stream. One of them checks if the comment is rude, while another identifies if this issue is about bug reproducibility. In this case, both DPRs need to conduct a filter to keep only issue creation events and parse out the \texttt{comments} field from the complex JSON structure. Additionally, the NLP procedures they rely on also have a lot of overlapping efforts (like word splitting). When these two DPRs are executed independently, the fluid ETL pipeline will run the same logic twice on the same data stream, causing significant resource waste.

To demonstrate the impact of this problem, we conduct an experiment to ingest GitHub data while applying both DPRs described above using manually written code at three different optimization levels. Specifically, we do not apply any optimizations in level 0, where two DPRs simply run successively. Optimization level 1 will let two DPRs share common efforts of identifying all issue creation events in the stream and parsing out their \texttt{comments} fields. For optimization level 2, we further examine the NLP procedures in both DPRs and eliminate higher-level redundant efforts, such as tokenization and named entity detection. As shown in Fig.~\ref{fig:fuse}, we can achieve a 5\% additional ingestion throughput by allowing two DPRs to share standard data operators (i.e., JSON parsing and filtering). When we carefully analyze the bespoke logic of two DPRs to further extract common logic for them to share, the ingestion throughput is more than doubled.

To maximize the utilization of any idle/cheap resources, it is natural for us to consider consolidating multiple DPRs into one so that redundant effort can be avoided during ingestion. Generally, if we frame two DPRs as functions $f_1(D) = x_1$ and $f_2(D) = x_2$, we are essentially creating a new function $f(D) = (x_1, x_2)$. Similar problems have been studied in a compiler context~\cite{gcse}, but the abstraction level they operate on (e.g., LLVM IR) is too low to
make any difference in fusing DPRs. A possible approach to solving this is to introduce a new, higher-level intermediate representation through the MLIR framework~\cite{mlir} and design customized optimization rules on top of it. If all DPRs involved are machine-generated, we can also leverage their shared code generation patterns to simplify the optimization procedure.

%% file: query.tex
\section{Query Planning with Partial Structures}
\label{sec:query}
As discussed in Section~\ref{sec:fetl}, structures built by fluid ETL pipelines are often \emph{partial} relative to ad-hoc queries due to resource availability and adaptive DPR management decisions. In such cases, classic query optimization techniques do not apply as they require leveraged structures to cover all query-related data to ensure correctness. Note that the concept of partial structures we introduced in this paper is fundamentally different from the partial index in classic literature~\cite{partial_index,partial2}.
Specifically, a partial index refers to an index built only on records that satisfy a predefined property (i.e., a user-defined filter), hence is \emph{partial to the whole table}. In contrast, partial structures defined in this paper are \emph{partial to data required to answer a specific query}. Classic query optimization techniques for partial indexes require the leveraged index to cover all query-related data so they cannot be applied for partial structures either.

\begin{figure}
	\centering
	\includegraphics[width=3.1in,trim=5mm 5mm 0 0]{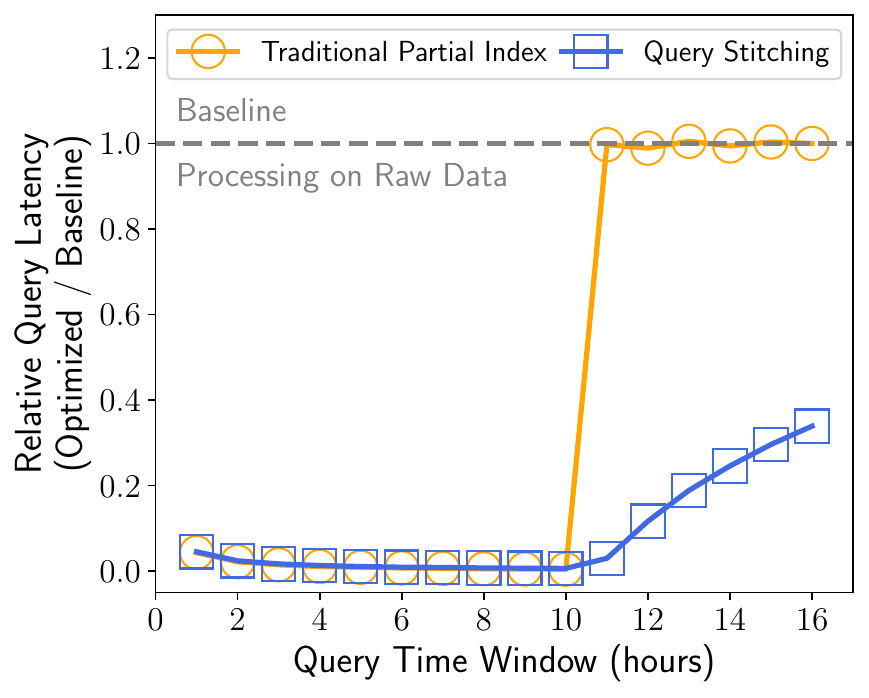}
	\vspace{-1mm}
	\caption{Effect of Query Stitching.}
	\label{fig:stitching}
	\vspace{-5mm}
\end{figure}

The key idea to address this problem is to decompose the underlying data into multiple partitions, rewrite the original query as subqueries operating on different partitions, and stitch their results together at the end. After decomposition, classic optimization techniques could utilize partial structures to accelerate some subqueries, providing a significant performance benefit. We call this idea \textbf{plan stitching}. To illustrate its power, we conduct a simple experiment on processing \textbf{(Q2)} described in Section~\ref{sec:fetl} with the presence of a partial index. Specifically, we issue \textbf{(Q2)} against events ingested from different time windows (e.g., in the last 2, 4, 6 hours). However, only the past 10 hours of ingested events
are covered by a hash index on \texttt{repo.id}. With traditional partial index optimization techniques,
the query engine can only leverage this partial index when \textbf{(Q2)} is conducted solely on data ingested
within the last 10 hours, but not those ingested further in the past. Nevertheless, for these cases, we
found that query stitching offers substantial benefits as shown in Fig.~\ref{fig:stitching}. Specifically, we decompose \textbf{(Q2)} into two parts, where one operates on data ingested over the past 10 hours using the hash index, and the other operates directly on the raw data remaining. Finally, we merge the results from these two parts.

In practice, we will face much more complicated scenarios. For a given ad-hoc query, there could be multiple partial structures available to accelerate some part of it. The coverage of these partial structures could be arbitrary and overlapping. When applying distinct decomposition strategies, we can leverage different combinations of partial structures in optimizing an ad-hoc query. Note that the coverage of partial structures built by a fluid ETL pipeline is continuous on the time dimension. Thus, for a query concerning $n$ records, there could be $2^n$ possible ways to decompose the underlying data. Enumerating through decomposition strategies is clearly impractical. Alternatively, if we decide to decompose the query region at partial structure coverage boundaries, simply combining the best plans optimized for different subqueries may well not be the best overall plan. Specifically, for two neighboring regions, leveraging the same partial structure would not introduce any stitching cost, whereas leveraging different partial structures may significantly reduce the cost of subqueries, thereby surpassing the stitching cost introduced. Lastly, since partial structures may be of different types (e.g., one is an index and the other is a small data cube), we may need to introduce non-trivial stitching operations in query rewriting to merge results from different subqueries. Hence, finding the best plan stitching strategy to fully leverage available partial structures remains an open research challenge.

% TODO we may need a figure for this?

%% file: optimize.tex
\section{Adaptive DPR Management}
\label{sec:adaptive}

In practice, many applications are running under resource or cost constraints. Generally, in a fluid ETL pipeline, every DPR we start can be viewed as a resource investment during ingestion, expecting the structures it builds to benefit some future queries. However, with limited idle/cheap resources at any time, we cannot have too many concurrent investments; otherwise, the ingestion system may not be able to provide enough ingestion throughput to keep up with data collection.
%Hence, when resource availability and the predicted workload changes, we will need to make dynamic decisions like terminating active DPRs to maintain high enough ingestion throughput or replacing/initiating new DPRs to match users' changing interests or when more resources become available.
% sentence is too long
Hence, when resource availability and predicted workload changes, we need to make dynamic decisions.
To keep up with data ingestion or to react to users' changing interests, we may need to replace, initialize, or terminate active DPRs.
To achieve this balance, we need to develop strategies for selecting the appropriate set of DPRs when the workload changes.

For better decision-making, we need a quantitative way to reason about the cost and benefit of introducing a DPR. As different types of resources (e.g., CPU cycles, disk bandwidth) can be bottlenecks for data ingestion, we need cost estimations for each DPR on all these resources. For an existing DPR, we can collect its basic execution statistics during ingestion as an estimation. For a newly generated one, we can run it against a small sample of the data stream to start a rough estimate and refine it later. For the benefit brought by a DPR for ad-hoc queries, we can leverage the cost model developed for query optimization.

With the cost and benefit estimation of DPRs in place, we can conduct an optimization to select the best DPRs to execute for a given query workload. The optimization should be constrained by current availability of idle/cheap resources, and should optimize for the best overall performance of the query workload. Once in a while, we will run this optimization on a new predicted workload, which can be approximated by blending past queries and predicted ones (given by the methods discussed in Section~\ref{sec:dprgen}). Target ingestion throughput should match the expected data collection rate, which can be derived by leveraging sequence-to-sequence learning methods. Note that the optimization described above can be treated as an instance of the knapsack problem~\cite{knapsack}, which is known to be NP-Complete. Thus, to reach a good selection within a reasonable amount of time, we might need to develop more efficient algorithms to approximate the optimal results. In addition to periodic reoptimization, a control theory-based method can provide a more incremental approach. Specifically, as workload changes, new DPRs are introduced for new query types, while those no longer serving any query type in the predicted workload are dropped. Meanwhile, the system will incrementally make adjustments when seeing performance fluctuation or available idle resources.

%% file: conc.tex
\section{Conclusion}
\label{sec:conc}
In this paper, we point out the excessive cost of existing solutions for providing ideal fresh data exploration. With a key insight, we envision a different approach that leveraging idle system capacity and/or cheap preemptive resources to run DPR speculatively to accelerate fresh data exploration workloads. In particular, we introduce a new type of ingestion system called fluid ETL pipelines that can start/stop arbitrary DPRs on demand without blocking ingestion. We verified the viability of our vision by simulating an exploration scenario over real-world datasets. Towards a comprehensive and practical solution, we identify several major challenges in DPR generation, query processing, and adaptive DPR management. We discussed these challenges in detail and outlined some potential directions to address them. In future work, we aim to develop core methodologies to address these new challenges and build a comprehensive and practical system to facilitate fresh data exploration.

% TODO remove thrifty, highlight the idle/cheap resources again.